\documentclass[%
 reprint, 
superscriptaddress, 
 amsmath, amssymb, 
 aps, 
]{revtex4-2}
\usepackage{subcaption}  
\usepackage{multirow}
\usepackage{graphicx}
\usepackage{dcolumn}
\usepackage{bm}
\usepackage{hyperref}

\begin{document}

\preprint{APS/123-QED}

\title{Surpassing the PLOB bound in continuous-variable quantum secret sharing \\ using a state-discrimination detector}

\author{Qin Liao}
\email{llqqlq@hnu.edu.cn}
\affiliation{College of Computer Science and Electronic Engineering, Hunan University, Changsha 410082, China}

\author{Chong Tang}
\affiliation{College of Computer Science and Electronic Engineering, Hunan University, Changsha 410082, China}

\author{Qingquan Peng}
\affiliation{Institute for Quantum Information $\&$ State Key Laboratory of High Performance Computing, College of Computer Science and Technology, National University of Defense Technology, Changsha 410073, China}

\author{Chao Ding}
\affiliation{School of Artificial Intelligence and Robotics, Hunan University, Changsha 410082, China}
\affiliation{Centre for Quantum Technologies, National University of Singapore, Singapore 117543, Singapore}

\author{Lingjin Zhu}
\affiliation{Hunan Institute of Metrology and Test, Changsha 410014, China}


\author{Yijun Wang}
\email{xxywyj@csu.edu.cn}
\affiliation{Center for Optoelectronic Information Engineering, School of Automation, Central South University, Changsha 410083, China}

\author{Xiquan Fu}
\affiliation{College of Computer Science and Electronic Engineering, Hunan University, Changsha 410082, China}

\date{\today}

\begin{abstract}

Continuous-variable quantum secret sharing (CVQSS) is a promising approach to ensuring multi-party information security. While CVQSS offers practical ease of implementation, its present performance remains limited. In this paper, we propose a novel CVQSS protocol integrated with a state-discrimination detector (SDD), dubbed SDD-CVQSS. In particular, we first develop the detailed procedure of SDD-CVQSS, which replaces the traditional coherent detector with an SDD and eliminates the long-standing necessary step of establishing multiple point-to-point quantum key distribution links between all users and the dealer. We then elaborate on the principle of the specifically designed SDD, which can efficiently discriminate mixed states with a much lower error probability. Finally, we construct a security model for SDD-CVQSS and derive its security bound against beam-splitting collective attacks. Numerical simulations show that SDD-CVQSS outperforms conventional CVQSS in both maximum transmission distance and secret key rate, even surpassing the PLOB bound. Additionally, we find that the performance degradation of SDD-CVQSS in long-distance transmission scenarios can be effectively compensated for using a post-selection scheme, providing a feasible way to achieve high-performance CVQSS.

\end{abstract}

\maketitle

\section{\label{sec:level1}INTRODUCTION}

A $(K, N)$ threshold secret sharing (SS) \cite{adi1979share} enables a legitimate user, called a dealer, to divide a string of secret keys into $n$ parts, which are distributed to a group of remote users through untrusted channels. Any single user cannot reconstruct the complete key information unless at least $K\;(K \leq N)$ users combine their shares. The security of SS mainly relies on the computational complexity of hard mathematical problems so that it can hardly be compromised by classical eavesdroppers within a very limited period \cite{blakley1979safeguarding}.

However, with the rapid development of quantum computing \cite{ladd2010quantum}, mathematical problems such as integer factorization \cite{grover1996fast} and discrete logarithm \cite{shor1994algorithms} can be efficiently solved by quantum algorithms \cite{brooks2019beyond}, seriously threatening the security of SS. 
To effectively cope with this threat, a quantum version of SS, namely quantum secret sharing (QSS), has been suggested as its security can be unconditionally guaranteed by the laws of quantum mechanics \cite{hillery1999quantum, karlsson1999quantum}. As one of the key techniques of QSS, continuous-variable (CV) QSS has become a research hotspot due to its simple signal preparation and strong compatibility with existing optical communication networks. 

The first CVQSS scheme was proposed by Ref. \cite{tyc2002share}, and Kogias et al. proved its theoretical security against both eavesdroppers and dishonest users \cite{kogias2017unconditional}. After that, Grice et al. showed that CVQSS can be implemented by sequentially sending weak coherent states which are more resilient to channel loss \cite{grice2019quantum}. Wu et al. subsequently extended this idea to thermal states, reducing the cost and complexity for implementing CVQSS \cite{wu2020passive}. To close the loopholes caused by imperfect devices, local local oscillator (LLO)-based CVQSS \cite{liao2024practical} and measurement-device-independent (MDI) CVQSS \cite{https://doi.org/10.1002/qute.202400505} were successively suggested to enhance the practical security of CVQSS system. Nevertheless, these CVQSS schemes are based on Gaussian modulation which is quite challenging to be implemented since current modulators are commonly limited by finite precision of sampling \cite{PhysRevA.86.032309}.  To tackle this issue, reference \cite{liao2021quantum} suggested a discretely-modulated (DM) CVQSS scheme where quadrature phase shift keying (QPSK) is applied instead of Gaussian modulation, reducing the difficulty of implementing CVQSS system. Subsequently, a multi-ring modulation strategy is developed to further enhance the performance of DM CVQSS \cite{liao2023continuous}.

The common point of all above-mentioned works is that the modulated coherent states, including Gaussian modulation and discrete modulation, are all measured by the receiver with a traditional coherent detector which cannot surpass the standard quantum limit (SQL) \cite{takeoka2008discrimination}. As a core theoretical benchmark in the field of quantum measurement, SQL defines the minimum error with which non-orthogonal states can be distinguished by direct measurement of the physical property of the light \cite{giovannetti2004quantum, chaves2013noisy}. In fact, quantum mechanics allows a lower error bound called Helstrom bound \cite{helstrom1969quantum} which can be approached or even achieved by optimizing discriminant strategy \cite{barnett1997experimental}. Tsujino et al. first confirmed that the SQL for discriminating two non-orthogonal states can be broken using near-unit-efficiency detector \cite{tsujino2011quantum}. After that, the SQLs for discriminating more than two non-orthogonal states were successively beaten by optimizing the test of multivariate hypothesis of coherent state events \cite{becerra2011m, muller2012quadrature}. Becerra et al. subsequently demonstrated the unconditional discrimination for QPSK-modulated coherent states (QMCSs) by using photon counting and adaptive measurements in the form of fast feedback \cite{becerra2013experimental}. Therefore, the performance of CVQSS would be further improved by appropriately replacing the traditional coherent detector with a certain kind of discriminant strategy. In fact, this idea was preliminarily verified by our previous work in terms of point-to-point continuous-variable quantum key distribution (CVQKD) system \cite{liao2018long}, and it exhibited excellent performance improvement in both binary-modulated CVQKD \cite{zhao2020security} and QPSK-modulated CVQKD \cite{zhao2024security}. 

Inspired by these previous works, we extend the idea from point-to-point quantum communication to multi-party quantum communication in this work, thus proposing a novel CVQSS protocol integrated with a specifically designed state-discrimination detector (SDD), which we call SDD-CVQSS. 
Specifically, we first develop the detailed procedure of SDD-CVQSS, which differs significantly from that of existing CVQSS \cite{grice2019quantum, wu2020passive, liao2024practical, https://doi.org/10.1002/qute.202400505, PhysRevA.86.032309, liao2021quantum, liao2023continuous}. The primary improvements of the SDD-CVQSS protocol lie in replacing the traditional coherent detector with an SDD and eliminating the long-standing step of establishing multiple point-to-point quantum key distribution links between all users and the dealer, which are beneficial for improving performance while reducing the complexity of the CVQSS system.
Then we design an SDD with multiple rounds of adaptive measurements, which is based on the maximum a posteriori probability (MAP) criterion. This detector is able to beat the SQL, enabling the dealer to efficiently discriminate the received mixed states with a substantially lower error probability.
We subsequently construct a security model for SDD-CVQSS by analyzing three distinct attack strategies and finally derive its security bound against the beam-splitting collective attacks.
Numerical simulations show that the proposed SDD-CVQSS outperforms conventional CVQSS in both maximum transmission distance and secret key rate, and its performance even surpasses the Pirandola-Laurenza-Ottaviani-Banchi (PLOB) bound \cite{pirandola2017fundamental}, which is a fundamental limit of repeaterless quantum communications. Moreover, we find that the performance of SDD-CVQSS can be further improved by increasing the number of adaptive measurement rounds, and its performance degradation in long-distance transmission scenarios can be effectively compensated for using a post-selection scheme.

This paper is structured as follows. In Sec. \ref{II}, we detail the proposed SDD-CVQSS protocol. In Sec. \ref{III}, we elaborate on the description of mixed states and the principle of the SDD. In Sec. \ref{IV}, the security analysis of SDD-CVQSS including the construction of security model and the calculation of the secret key rate is presented. Performance analysis and discussion are provided in Sec. \ref{V}, and conclusions are drawn in Sec. \ref{VI}.

\section{SDD-CVQSS protocol}\label{II}

In conventional CVQSS, a traditional coherent detector, i.e., a heterodyne detector, is usually adopted at dealer's side for measuring mixed coherent states. While in our proposed SDD-CVQSS, the detection strategy is quite different by replacing the coherent detector with SDD. 
For simplicity, here we elaborate on the SDD-CVQSS protocol by detailing the basic $(2, 2)$ threshold scheme of SDD-CVQSS, its more complicated $(N, N)$ threshold scheme can be derived with similar idea.

\begin{figure*}
\includegraphics[width = \textwidth]{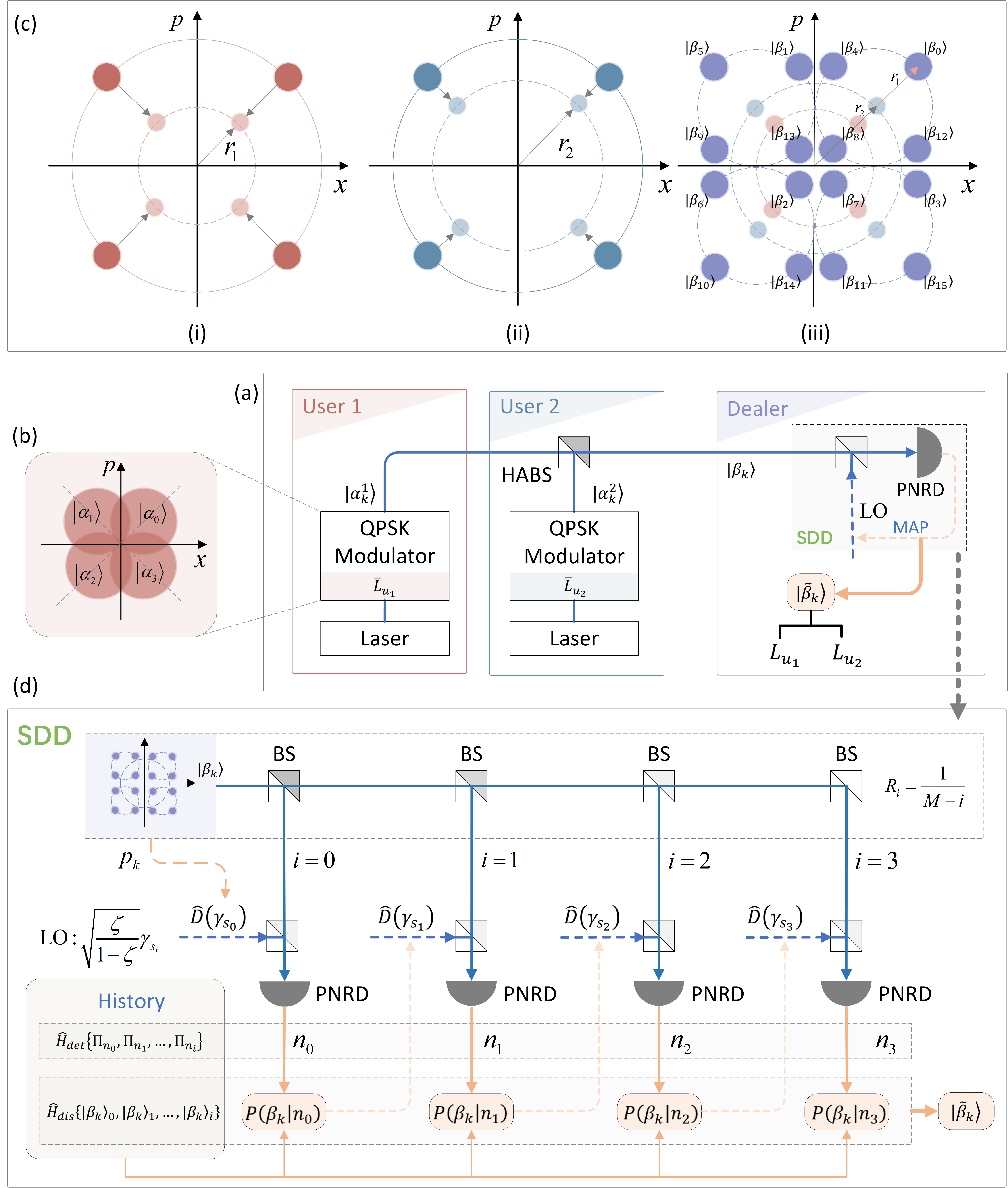}
\caption{(a) Schematic diagram of a $(2,2)$ threshold SDD-CVQSS scheme, where two remote users (user 1 and user 2) are sequentially connected to the dealer via an untrusted quantum channel. (b) Description of QPSK with a constellation of four coherent states. Four QMCSs are located in four quadrants with modulated phase of $\{\pi/4$, $3\pi/4$, $5\pi/4$, $7\pi/4\}$, respectively. (c) Forming process of mixed state. (i) QMCSs generated by user 1 (red solid balls) are attenuated (red transparency balls) after being transmitted from user 1 to the dealer. (ii) QMCSs generated by user 2 (blue solid balls) are attenuated (blue transparency balls) after being transmitted from user 2 to the dealer. (iii) The constellation of the actual mixed state received at the dealer's side (purple solid balls). The radius $r$ is inversely correlated with the transmission distance between user and the dealer. (d) Schematic diagram of SDD for discriminating mixed states with $M$ adaptive measurements.}
\label{fig:SDDF}
\end{figure*}

The basic $(2, 2)$ threshold scheme of SDD-CVQSS is shown in Fig. \ref{fig:SDDF}(a) in which two remote users (user 1 and user 2) are orderly connected with the dealer through an untrusted quantum channel. Each user locally prepares QMCSs and sends them to the dealer who measures the incoming signals with SDD. The main steps of SDD-CVQSS are described below.

\textit{Step 1}. An $m$-bit binary sequence $\bar{L}_{u_1}$ is first generated by user 1, and this sequence is then divided into pairs and mapped onto four coherent states according to the mapping rule $00\rightarrow|\alpha_0\rangle, 01\rightarrow|\alpha_1\rangle, 10\rightarrow|\alpha_2\rangle, 11\rightarrow|\alpha_3\rangle$, where $\alpha$ is the coherent amplitude. As a result, the QMCSs, i.e., \{$\left|\alpha_{k_1}^{1}\right\rangle\}=\{\left|\alpha e^{i(2 k_1+1) \pi / 4}\right\rangle$\} where $k_{1}\in\{0,1,2,3\}$, can be prepared, and user 1 successively sends them to user 2.

\textit{Step 2}. Similarly, user 2 also randomly generates another $m$-bit binary sequence $\bar{L}_{u_2}$ and independently prepares QMCSs according to each pair that is divided from this sequence. By following the same mapping rule, user 2's prepared QMCSs can be expressed as \{$\left|\alpha_{k_2}^{2}\right\rangle\}=\{\left|\alpha e^{i(2 k_2+1) \pi / 4}\right\rangle$\} where $k_{2}\in\{0,1,2,3\}$, which are successively coupled to the same spatiotemporal mode as the incoming QMCSs prepared by user 1 through a highly asymmetric beam splitter (HABS). The mixed signals are then sent to the dealer.

\textit{Step 3}. After passing the untrusted quantum channel, the mixed signals \{$|\beta_{k}\rangle\}=\{|\sqrt{\eta_{1}}\alpha_{k_1}^{1}\rangle+|\sqrt{\eta_{2}}\alpha_{k_2}^{2}\rangle\}, k=4k_1+k_2$ are successively measured by the dealer using an SDD with MAP criterion, where $\eta_1$ and $\eta_2$ denote the channel transmittances experienced by the signals between user 1 and the dealer, and between user 2 and the dealer, respectively. As a result, the estimated states, denoted by \{$|\widetilde{\beta}_{k}\rangle$\}, can be obtained by the dealer.

\textit{Step 4}. By following the reverse mapping rule, the dealer re-maps these estimated states into a binary sequence $L=l_{1}l_{2}\ldots l_{2m-1}l_{2m}$ which can be further divided into $L_{u_{1}}=l_{1}l_{2}...l_{a_{m-1}}l_{a_{m}}$ and $L_{u_{2}}=l_{3}l_{4}...l_{b_{m-1}}l_{b_{m}}$, where 
\begin{equation}
\begin{split}
a_m=\begin{cases}4\left\lfloor\frac{m-1}{2}\right\rfloor+1\, m\, mod\, 2=1\\4\left\lfloor\frac{m-1}{2}\right\rfloor+2\, m\, mod\, 2=0\end{cases} \label{eq:1}
\end{split}
\end{equation}
and
\begin{equation}
\begin{split}
b_{m}=\begin{cases}4\left\lfloor\frac{m-1}{2}\right\rfloor+3\, m\, mod\, 2=1\\4\left\lfloor\frac{m-1}{2}\right\rfloor+4\, m\, mod\, 2=0\end{cases}, \label{eq:2}
\end{split}
\end{equation}
where $\lfloor\cdot\rfloor$ is the downward integer symbol. By doing this, the dealer and each user hold sufficiently related raw keys, i.e., ${L}_{u_1}$ and $\bar{L}_{u_1}$ for the dealer and user 1, and ${L}_{u_2}$ and $\bar{L}_{u_2}$ for the dealer and user 2.

\textit{Step 5}. Each user and the dealer perform one-way error correction and privacy amplification \cite{zhao2024security} on their related raw keys to extract point-to-point secret keys $K_{1}=l_{1}l_{2}\ldots l_{a_{t-1}}l_{a_{t}}$ and $K_{2}=l_{3}l_{4}...l_{b_{r-1}}l_{b_r},\,$ where $t,\,r < m$. Finally, the dealer encodes a message $Y$ as $E=Y\oplus K_{1}\oplus K_{2}$. Obviously, only if user 1 and user 2 combine their shares can this encrypted message $E$ be correctly decoded.

\section{Principle of State-Discrimination Detector}\label{III}

In the proposed SDD-CVQSS protocol, an SDD is employed at the dealer's side to measure the incoming states, which is entirely different from conventional CVQSS. In what follows, we first describe the incoming mixed states and then provide a detailed explanation of how the SDD discriminates these mixed states.

\subsection{\label{sec:MSD}Mixed state description}

In \textit{Step 1} and \textit{Step 2}, user 1 and user 2 independently prepare QMCS according to their randomly generated binary sequences. The discrete modulation strategy for QMCS requires four non-orthogonal coherent states and its presentation in phase space is depicted in Fig. \ref{fig:SDDF}(b). The output QMCS for each user can be respectively expressed by 
\begin{equation}
\rho_1=\frac{1}{4}\sum_{k=0}^3|\alpha_k^{1}\rangle\langle\alpha_k^{1}|
\end{equation}
and
\begin{equation}
\rho_2=\frac{1}{4}\sum_{k=0}^3|\alpha_k^{2}\rangle\langle\alpha_k^{2}|.
\end{equation}
After the above two QMCSs couple into the same spatiotemporal mode and pass through the quantum channel, the mixed states can be depicted as 
\begin{equation}
\rho_{m}=\frac{1}{16}\sum_{i=0}^3\sum_{j=0}^3(\eta_{1}|\alpha_i^{1}\rangle\langle\alpha_i^{1}|\otimes\eta_{2}|\alpha_j^{2}\rangle\langle\alpha_j^{2}|),
\end{equation}
whose forming process is illustrated in Fig. \ref{fig:SDDF}(c). Specifically, after passing the quantum channel, QMCSs generated by user 1 and user 2 are attenuated to different degrees ($r_1 < r_2$) due to the different transmission distances, so that the actual mixed state received by the dealer may belong to one of the 16 possible states, which is described in Table \ref{tab:table2}.

\begin{table*}
\scriptsize
\centering
\caption{\label{tab:table2} 16 possible mixed states that the dealer may receive.}
\resizebox{0.7\textwidth}{!}{%
    
    \begin{tabular}{c c ccc} 
    \hline\hline
    user 1  & user 2 & mixed state & amplitude & phase  \\ 
    \hline
    \multirow{4}{*}{00} & 00 & $|\beta_0\rangle=|\sqrt{\eta_{1}}\alpha_{0}^{1}\rangle+|\sqrt{\eta_{2}}\alpha_{0}^{2}\rangle$ & $\eta_1\alpha^2+\eta_2\alpha^2$ & $\pi/{4}$  \\
                       & 01 & $|\beta_1\rangle=|\sqrt{\eta_{1}}\alpha_{0}^{1}\rangle+|\sqrt{\eta_{2}}\alpha_{1}^{2}\rangle$ & $\sqrt{(\eta_{1}\alpha^{2})^{2}+(\eta_{2}\alpha^{2})^{2}}$ & $\pi/{4}+\arctan\left(\eta_{2}/\eta_{1}\right)$  \\
                       & 10 & $|\beta_2\rangle=|\sqrt{\eta_{1}}\alpha_{0}^{1}\rangle+|\sqrt{\eta_{2}}\alpha_{3}^{2}\rangle$ & $\eta_2\alpha^2-\eta_1\alpha^2$ & $5\pi/{4}$  \\
                       & 11 & $|\beta_3\rangle=|\sqrt{\eta_{1}}\alpha_{0}^{1}\rangle+|\sqrt{\eta_{2}}\alpha_{4}^{2}\rangle$ & $\sqrt{(\eta_{1}\alpha^{2})^{2}+(\eta_{2}\alpha^{2})}$ & $7\pi/{4}+\arctan\left(\eta_{1}/\eta_{2}\right)$  \\ 
    \hline
    \multirow{4}{*}{01} & 00 & $|\beta_4\rangle=|\sqrt{\eta_{1}}\alpha_{1}^{1}\rangle+|\sqrt{\eta_{2}}\alpha_{0}^{2}\rangle$ & $\sqrt{(\eta_{1}\alpha^{2})^{2}+(\eta_{2}\alpha^{2})^{2}}$ & $\pi/{4}+\arctan\left(\eta_{1}/\eta_{2}\right)$  \\
                       & 01 & $|\beta_5\rangle=|\sqrt{\eta_{1}}\alpha_{1}^{1}\rangle+|\sqrt{\eta_{2}}\alpha_{1}^{2}\rangle$ & $\eta_1\alpha^2+\eta_2\alpha^2$ & $3\pi/{4}$  \\
                       & 10 & $|\beta_6\rangle=|\sqrt{\eta_{1}}\alpha_{1}^{1}\rangle+|\sqrt{\eta_{2}}\alpha_{2}^{2}\rangle$ & $\sqrt{(\eta_{1}\alpha^{2})^{2}+(\eta_{2}\alpha^{2})^{2}}$ & $3\pi/{4}+\arctan\left(\eta_{2}/\eta_{1}\right)$  \\
                       & 11 & $|\beta_7\rangle=|\sqrt{\eta_{1}}\alpha_{1}^{1}\rangle+|\sqrt{\eta_{2}}\alpha_{3}^{2}\rangle$ & $\eta_2\alpha^2-\eta_1\alpha^2$ & $7\pi/{4}$  \\ 
    \hline
    \multirow{4}{*}{10} & 00 & $|\beta_8\rangle=|\sqrt{\eta_{1}}\alpha_{2}^{1}\rangle+|\sqrt{\eta_{2}}\alpha_{0}^{2}\rangle$ & $\eta_2\alpha^2-\eta_1\alpha^2$ & $\pi/{4}$  \\
                       & 01 & $|\beta_9\rangle=|\sqrt{\eta_{1}}\alpha_{2}^{1}\rangle+|\sqrt{\eta_{2}}\alpha_{1}^{2}\rangle$ & $\sqrt{(\eta_{1}\alpha^{2})^{2}+(\eta_{2}\alpha^{2})^{2}}$ & $3\pi/{4}+\arctan\left(\eta_{1}/\eta_{2}\right)$  \\
                       & 10 & $|\beta_{10}\rangle=|\sqrt{\eta_{1}}\alpha_{2}^{1}\rangle+|\sqrt{\eta_{2}}\alpha_{2}^{2}\rangle$ & $\eta_1\alpha^2+\eta_2\alpha^2$ & $5\pi/{4}$  \\
                       & 11 &                  $|\beta_{11}\rangle=|\sqrt{\eta_{1}}\alpha_{2}^{1}\rangle+|\sqrt{\eta_{2}}\alpha_{3}^{2}\rangle$ & $\sqrt{(\eta_{1}\alpha^{2})^{2}+(\eta_{2}\alpha^{2})^{2}}$ & $5\pi/{4}+\arctan\left(\eta_{2}/\eta_{1}\right)$  \\
    \hline
    \multirow{4}{*}{11} & 00 &               $|\beta_{12}\rangle=|\sqrt{\eta_{1}}\alpha_{3}^{1}\rangle+|\sqrt{\eta_{2}}\alpha_{0}^{2}\rangle$ & $\sqrt{(\eta_{1}\alpha^{2})^{2}+(\eta_{2}\alpha^{2})^{2}}$ & $\pi/{4}-\arctan\left(\eta_{1}/\eta_{2}\right)$  \\ 
                       & 01 & $|\beta_{13}\rangle=|\sqrt{\eta_{1}}\alpha_{3}^{1}\rangle+|\sqrt{\eta_{2}}\alpha_{1}^{2}\rangle$ & $\eta_2\alpha^2-\eta_1\alpha^2$ & $3\pi/{4}$  \\
                       & 10 & $|\beta_{14}\rangle=|\sqrt{\eta_{1}}\alpha_{3}^{1}\rangle+|\sqrt{\eta_{2}}\alpha_{2}^{2}\rangle$ & $\sqrt{(\eta_{1}\alpha^{2})^{2}+(\eta_{2}\alpha^{2})^{2}}$ & $5\pi/{4}+\arctan\left(\eta_{1}/\eta_{2}\right)$  \\
                       & 11 & $|\beta_{15}\rangle=|\sqrt{\eta_{1}}\alpha_{3}^{1}\rangle+|\sqrt{\eta_{2}}\alpha_{3}^{2}\rangle$ & $\eta_1\alpha^2+\eta_2\alpha^2$ & $7\pi/{4}$  \\
    \hline\hline
    \end{tabular}
}
\end{table*}

\subsection{\label{sec:DSDD} Design of state discrimination detector}

In \textit{Step 3}, the incoming mixed signal is measured by the dealer using SDD illustrated in Fig. \ref{fig:SDDF}(d). The SDD, which consists of $M$ adaptive measurements, first divides the mixed signal into $M$ equal-intensity branches by $M$ beam splitters whose reflectivities are $R_i=\frac{1}{M-i}$, where $i=0, 1, ..., M-1$. For each branch, the splitting signal $\frac{1}{\sqrt{M}}|\beta_k\rangle$ is subsequently displaced by a displacement operator $\widehat{D}(\gamma_{s_{i}})$ which is achieved by interfering the splitting signal with a local oscillator (LO) via another beam splitter of transmittance $\zeta \rightarrow 1$ \cite{9311620}. Assuming that $|\beta_k\rangle$ is selected with the prior probability $p_k$, the coherent amplitude of LO can be set to $\sqrt{\frac{\zeta}{1-\zeta}}\gamma_{s_i}$ where $\gamma_{s_i}=\frac{1}{\sqrt{M}}\beta_{k_i}$, so that the splitting signal can be displaced as 

\begin{equation}
\begin{split}
\widehat{D}\left(\gamma_{s_{i}}\right)\left|\frac{\beta_{k}}{\sqrt{M}}\right\rangle&=\left|\sqrt{\frac{\zeta}{M}}\beta_{k}+\sqrt{1-\zeta}\sqrt{\frac{\zeta}{1-\zeta}}\gamma_{s_{i}}\right\rangle
\\&=\left|\sqrt{\zeta}\left(\frac{\beta_{k}}{\sqrt{M}}+\gamma_{s_{i}}\right)\right\rangle. \label{eq:11D}
\end{split}
\end{equation}

By taking into account the influence of the thermal noise which is caused by the rising temperature of the load resistor when SDD continuously works \cite{PhysRev.130.2529}, the density operator of the signal after displacement can be expressed as \cite{PhysRev.131.2766}
\begin{align}
\widehat{\rho}_{th}(\beta_{k}, \gamma_{s_{i}})=\frac{1}{\pi N_{t}}\int_{\mathbb{C}}e^{-\frac{|\tau|^{2}}{N_{t}}}\left|\sqrt{\zeta}(\frac{\beta_{k}}{\sqrt{M}}-\gamma_{s_{i}})+\tau\right\rangle
\nonumber\\ \left\langle\sqrt{\zeta}(\frac{\beta_{k}}{\sqrt{M}}-\gamma_{s_{i}})+\tau\right|\mathrm{d}^{2}\tau, \label{eq:4}
\end{align}
where $N_t$ denotes the average number of thermal photons and characterizes the thermal noise level of SDD. 

Subsequently, a photon number resolving detector (PNRD) is adopted to measure the displaced state. The complete quantum mechanical description of this measurement in the form of positive operator-valued measure (POVM) can be given by
\begin{align}
\widehat{\Pi}_{{n_i}}=\sum_{j={n_i}}^{\infty}(_{{n_i}}^{j})\eta_{s}^{{n_i}}(1-\eta_{s})^{j-{n_i}}|j\rangle\langle j|, \label{eq:5}
\end{align}
where $\eta_{s}$ is the quantum efficiency of the PNRD and $n_i$ is the detection photon number on $i$-th branch. Apparently, $\Pi_0$ will click if the hypothesis is correct, which means that the input field of this branch will be displaced to vacuum so that the PNRD cannot detect any photon.

The conditional probability of detecting photon number $n_i$ at the PNRD can be obtained by
\begin{align}
&P(n_i|\beta_{k},\gamma_{s_{i}})=\mathrm{Tr}(\hat{\Pi}_{n_i}\hat{\rho}_{th}(\beta_{k},\gamma_{s_{i}}))
\nonumber \\&=\frac{(\eta_{s}N_{t})^{n_i}}{(\eta_{s}N_{t}+1)^{n_i+1}}e^{-\frac{\zeta\overline{N}+ \nu}{N_{t}+1/\eta_{s}}}\times L_{n_i}\left(-\frac{\zeta\overline{N}+ \nu}{N_{t}(\eta_{s}N_{t}+1)}\right),  \label{eq:6}
\end{align}
where $L_{n_i}(\cdot)$ is the Laguerre polynomial of order $n_i$ and the average number of photons $\overline{N}$ can be calculated by
\begin{equation}
\begin{split}
  \overline{N} &=\frac{|\beta_{k}|^{2}}{4} + |\gamma_{s_{i}}|^{2} - \xi \left| \beta_{k} \right| |\gamma_{s_{i}}| 
  \\ &\times \cos \left( \arg(\beta_{k}) - \arg(\gamma_{s_{i}}) \right), 
\end{split}
\end{equation}
where $\xi\in[0, 1]$ is the interference visibility caused by active stabilization
of the input power and relative phase difference between the input signal and the LO, which can be obtained from the interference measurement \cite{becerra2015photon}.

After measurement, the posterior probabilities of all possible states (i.e., \{$|\beta_k\rangle$\}) can be derived using Bayesian inference based on current detection history $\widehat{H}_{det}=\{\Pi_{n_0},\Pi_{n_1},...,\Pi_{n_i}\}$ and displacement history $\widehat{H}_{dis}=\{ |\beta_k\rangle _1, |\beta_k\rangle _2,...,|\beta_k\rangle _i\}$. For each possible state $|\beta_k\rangle$ in $i$-th adaptive measurement, its posterior probability can be obtained as
\begin{align}
P(\beta_{k}|n_0,...,n_{i})=\frac{P(\beta_{k}|n_0,...,n_{i-1})P(n_{i}|\beta_{k},\gamma_{s_{i}})}{\sum_{k=0}^{15}P(\beta_{k}|n_0,...,n_{i-1})P(n_{i}|\beta_{k},\gamma_{s_{i}})}, 
\label{eq:post}
\end{align}
where $P(\beta_{k}|n_0,...,n_{i-1})$ is the posterior probability in $(i-1)$-th adaptive measurement. According to the MAP criterion, the possible state with the highest posterior probability 
\begin{equation}
|\gamma_{s_{i+1}}\rangle=\mathrm{argmax}_{|\beta_{k}\rangle}P(\beta_{k}|\gamma_{s_{i}}, n_{i})
\end{equation}
will be selected as an input hypothesis for the next adaptive measurement. Note that the probabilities of all possible states are dynamically updated and the posterior probabilities of the states will become the prior probabilities of the states in the next adaptive measurement. This is because the latest measurement result and the selected state of $i$-th adaptive measurement will be respectively added into histories $\widehat{H}_{det}$ and $\widehat{H}_{dis}$.

By recursively calculating the Eq. (\ref{eq:post}), the posterior probability for each possible state $|\beta_k\rangle$ after $M$ adaptive measurements can be obtained as
\begin{align}
P(\beta_{k}|n_0,...,n_{M-1})=\frac{p_{k}P(n_{0}|\beta_{k}, \gamma_{s_{0}})\prod_{i=1}^{M-1}P(n_{i}|\beta_{k}, \gamma_{s_{i}})}{\sum_{k=0}^{15}p_{k}P(n_{0}|\beta_{k}, \gamma_{s_{0}})\prod_{i=1}^{M-1}P(n_{i}|\beta_{k}, \gamma_{s_{i}})}. \label{eq:ppost}
\end{align}
See Appendix \ref{app:1} for its detailed derivation. Finally, the estimated state $|\widetilde{\beta}_{k}\rangle$ of SDD corresponds to the one with the maximum of the posterior probabilities in $M$-th adaptive measurement, which is decided as
\begin{equation}
|\tilde{\beta_k}\rangle=\mathrm{argmax}_{|\beta_{k}\rangle}p_{k}P(n_{0}|\beta_{k}, \gamma_{s_{0}})\prod_{i=1}^{M-1}P(n_{i}|\beta_{k}, \gamma_{s_{i}}). \label{eq:est_beta}
\end{equation}

\section{Security analysis}\label{IV}

In this section, we first construct the security model of SDD-CVQSS and then derive its calculation of the secret key rate. 

\subsection{\label{sec:securityModel} Security model}

\begin{figure}
\includegraphics[width = \columnwidth]{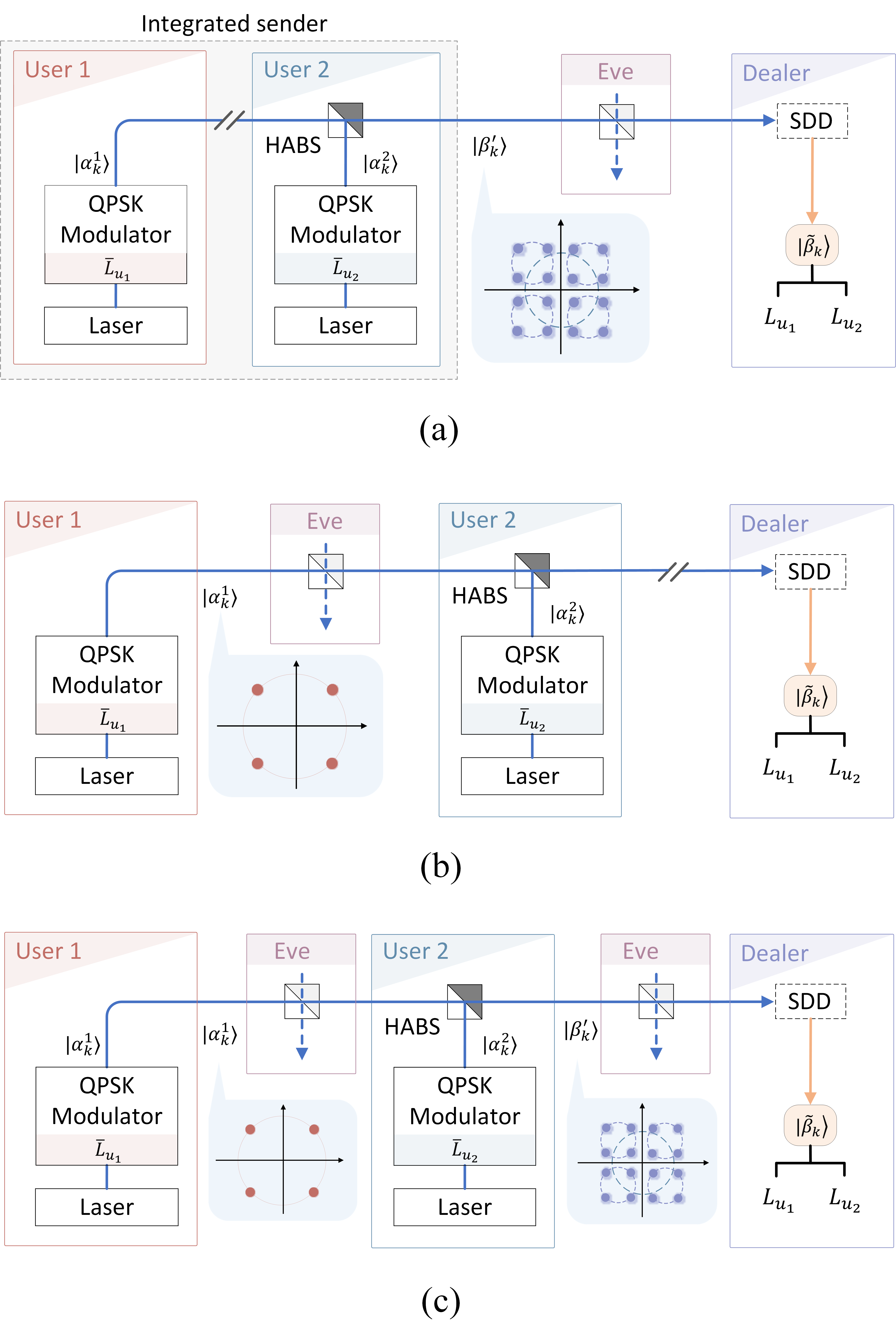}
\caption{Security models of SDD-CVQSS. (a) User 1 and user 2 are deemed an integrated sender, who prepares quantum states using 16QAM-like modulation strategy. (b) Eve only launches the attack between user 1 and user 2. (c) Eve launches the attack not only between user 1 and user 2 but also between user 2 and the dealer.}
\label{fig:system_excess}
\end{figure}

As described in Sec.\ref{II}, user 1 first prepares QMCSs and sends them to user 2 who also independently prepares QMCSs with the same QPSK modulation strategy, so that the mixed states at user 2's output can be expressed as $\{|\beta'_k\rangle\}=\{|\sqrt{\eta_1/\eta_2}\alpha^1_{k_1}+\alpha^2_{k_2}\rangle\}$. By regarding user 1 and user 2 as an integrated sender, the security model of SDD-CVQSS can be deemed point-to-point communication link with 16QAM-like modulation strategy shown in Fig.\ref{fig:system_excess}(a). Assuming the quantum channel between the sender and the receiver is a photon loss channel and the eavesdropper Eve applies the beam-splitting collective attack to obtain information about the secret key, the mixed state is split into
\begin{equation}
\begin{split}
|\beta'_k\rangle &\rightarrow \sqrt{\eta_2}|\beta'_k\rangle_D \otimes \sqrt{1-\eta_2}|\beta'_k\rangle_E \\
&=|\sqrt{\eta_1}\alpha^1_{k_1}+\sqrt{\eta_2}\alpha^2_{k_2}\rangle_D \\
&\otimes \left|\sqrt{\frac{(1-\eta_2)\eta_1}{\eta_2}}\alpha^1_{k_1}+\sqrt{1-\eta_2}\alpha^2_{k_2}\right\rangle_E. 
\label{eq:beta'}
\end{split}
\end{equation}

This splitting process reflects the information leakage caused by Eve's attack, where the state $|\beta_k\rangle=|\sqrt{\eta_1}\alpha^1_{k_1}+\sqrt{\eta_2}\alpha^2_{k_2}\rangle_D$ is received by the dealer and the state $|\varepsilon_k\rangle=|\sqrt{(1-\eta_2)\eta_1/\eta_2}\alpha^1_{k_1}+\sqrt{1-\eta_2}\alpha^2_{k_2}\rangle_E$ is intercepted by Eve. The received state $|\beta_k\rangle$ is then estimated by the SDD, and the estimated state $|\widetilde{\beta}_{k}\rangle$ can be finally obtained by Eq. (\ref{eq:est_beta}).

It is worthy of note that apart from the above attack strategy, Eve can also eavesdrop on SDD-CVQSS by intercepting QMCSs transmitted between user 1 and user 2, given that the quantum channel between these two users is untrusted. This gives rise to two additional security models, i.e., Eve only launches the attack between user 1 and user 2, as shown in Fig. \ref{fig:system_excess}(b), and Eve launches the attack not only between user 1 and user 2 but also between user 2 and the dealer, as shown in Fig. \ref{fig:system_excess}(c). For the security model (b), Eve's attack is apparently limited because she can only obtain information related to user 1, thereby leading to an underestimation of the information accessible to Eve. For the security model (c), it is actually equivalent to the security model (a) because it suffices for Eve to obtain information by intercepting $|\beta'_k\rangle$, which contains all users' modulated information. In other words,  if Eve intercepts $|\beta'_k\rangle$, she can directly infer user 1's modulated information without intercepting user 1's QMCSs between user 1 and user 2. This implies that Eve's attack between user 1 and user 2 is redundant for acquiring additional useful information. Therefore, we calculate the secret key rate of SDD-CVQSS based on the security model (a), which is detailed in next subsection.

\subsection{\label{sec:SKR} Calculation of the secret key rate}

The asymptotic secret key rate of SDD-CVQSS under the beam-splitting collective attack is bounded by 
\begin{equation}
R \geq I_{UD}-\chi,
\label{eq:skr}
\end{equation}
where $I_{UD}$ is the mutual information between the integrated sender and the dealer, and $\chi$ is the Holevo quantity of Eve's state, which specifically quantifies the maximum amount of information Eve can extract from the transmitted signal. 

Specifically, the mutual information $I_{UD}$ is given as
\begin{equation}
I_{UD}=H(U)-H(U|D), 
\label{eq:idu1}
\end{equation}
where $H(U)=-\sum_{k=0}^{15}p_{k}\mathrm{log}_{2}p_{k}$ is the entropy of the transmitted signal by the integrated sender, and the conditional entropy $H(U|D)$ quantifies the remaining uncertainty concerning the transmitted signal $|\beta'_k\rangle$ conditioned on the dealer’s estimated signal $|\tilde{\beta}\rangle$. Since the transmitted signal $|\beta'_k\rangle$ follows the same probability distribution as the received signal $|\beta_k\rangle$ of dealer's side, we have $P(\beta'_{k}|n_{0}, ... , n_{M-1})=P(\beta_{k}|n_{0}, ... , n_{M-1})$. Therefore, $H(U|D)$ can be calculated as
\begin{equation}
\begin{split}
H(U|D)&=\sum_{n_{0}=0}^{\infty}... \sum_{n_{M-1}=0}^{\infty}p(n_{0}, ... , n_{M-1})H(U|n_{0}, ... , n_{M-1})\\&=-\sum_{n_{0}=0}^{\infty}... \sum_{n_{M-1}=0}^{\infty}p(n_{0}, ... , n_{M-1})\\&\times\sum_{k=0}^{15}P(\beta_{k}|n_{0}, ... , n_{M-1})\mathrm{log}_{2}P(\beta_{k}|n_{0}, ... , n_{M-1}).
\label{eq:u1d}
\end{split}
\end{equation}

Based on the fact $p(n_{0}, ... , n_{M-1})P(\beta_{k}|n_{0}, ... , n_{M-1})$ $=P(n_{0}, ... , n_{M-1}|\beta_{k})$, and the equation
\begin{equation}
\begin{split}
P(n_{0}, ... , n_{M-1}|\beta_{k})=P(n_{0}|\beta_{k}, \gamma_{s_{0}})\prod_{i=1}^{M-1}P(n_{i}|\beta_{k}, \gamma_{s_{i}}),
\end{split}
\end{equation}
Eq. (\ref{eq:u1d}) can be further rewritten as
\begin{equation}
\begin{split}
H(U|D)&=-\sum_{n_{0}=0}^{\infty}... \sum_{n_{M-1}=0}^{\infty} \sum_{k=0}^{15}p_kP(n_{0}, ... , n_{M-1}|\beta_{k})\\&\times\mathrm{log}_{2}P(\beta_{k}|n_{0}, ... , n_{M-1})\\&=-\sum_{n_{0}=0}^{\infty}... \sum_{n_{M-1}=0}^{\infty} \sum_{k=0}^{15}p_kP(n_{0}|\beta_{k}, \gamma_{s_{0}})\\&\prod_{i=1}^{M-1}P(n_{i}|\beta_{k}, \gamma_{s_{i}})\mathrm{log}_{2}P(\beta_{k}|n_{0}, ... , n_{M-1}).
\label{eq:HUD}
\end{split}
\end{equation}

Given that the displacement operation of each adaptive measurement depends solely on the immediately preceding photon number detection result, the conditional probabilities $P(\beta_{k}|n_{0}, \ldots, n_{M-1})$ and $P(\beta_{k}|n_{M-1})$ can be considered equivalent. This equivalence arises because the adaptive measurement process forms a Markov chain \cite{10.1093/biomet/57.1.97}, where the current estimated state $|\beta_{k}\rangle$ is determined entirely by the most recent measurement $n_{M-1}$, rendering all prior measurements $n_0, \ldots, n_{M-2}$ conditionally independent. Therefore, the explicit expression for the mutual information $I_{UD}$ can be derived by using Eq. (\ref{eq:ppost}) and Eq. (\ref{eq:HUD}), that is
\begin{widetext}
\begin{equation}
\begin{split}
I_{UD}&=-\sum_{k=0}^{15}p_{k}\log_{2}p_k-\left(-\sum_{n_{0}=0}^{\infty}... \sum_{n_{M-1}=0}^{\infty} \sum_{k=0}^{15}p_kP(n_{0}|\beta_{k}, \gamma_{s_{0}})\prod_{i=1}^{M-1}P(n_{i}|\beta_{k}, \gamma_{s_{i}})\mathrm{log}_{2}P(\beta_{k}| n_{M-1})\right)\\&=-\sum_{k=0}^{15}p_{k}\left(\log_{2}p_k-\sum_{n_{0}=0}^{\infty}. . . \sum_{n_{M-1}=0}^{\infty}P(n_{0}|\beta_{k}, \gamma_{s_{0}})\prod_{i=1}^{M-1}P(n_{i}|\beta_{k}, \gamma_{s_{i}})\right.\\&\quad\left.\times\log_{2}\frac{p_{k}P(n_{0}|\beta_{k}, \gamma_{s_{0}})\prod_{i=1}^{M-1}P(n_{i}|\beta_{k}, \gamma_{s_{i}})}{\sum_{k=0}^{15}p_{k}P(n_{0}|\beta_{k}, \gamma_{s_{0}})\prod_{i=1}^{M-1}P(n_{i}|\beta_{k}, \gamma_{s_{i}})}\right). 
\label{eq:IIUD}
\end{split}
\end{equation}
\end{widetext}

For Eve, she can obtain information by accessing random state which takes its value from the set $\{\hat{\rho}_{E|\beta_k}=|\varepsilon_{k}\rangle\langle\varepsilon_{k}|\}$ with a prior probability $p_k$. Then the mixed state eavesdropped by Eve can be represented by the density operator
\begin{equation}
\begin{split}
\hat{\rho}_{E}&=\sum_{k=0}^{15}p_{k}\hat{\rho}_{E|\beta_k}
=\sum_{k=0}^{15}p_{k}|\varepsilon_{k}\rangle\langle\varepsilon_{k}|\\&=\sum_{k=0}^{15}p_{k} \left|\sqrt{\frac{(1-\eta_2)\eta_1}{\eta_2}}\alpha^1_{k_1}+\sqrt{1-\eta_2}\alpha^2_{k_2}\right\rangle \\& \left\langle\sqrt{\frac{(1-\eta_2)\eta_1}{\eta_2}}\alpha^1_{k_1}+\sqrt{1-\eta_2}\alpha^2_{k_2}\right|. 
\label{eq:r}
\end{split}
\end{equation}

Eve subsequently performs a POVM on state $\hat{\rho}_{E}$. According to Holevo theorem, if direct reconciliation (DR) is applied, Eve's information about the integrated sender's state is bounded by the Holevo bound $\chi_{UE}$ as
\begin{equation}
\begin{split}
\chi\rightarrow\chi_{UE} = S(\hat{\rho}_E) - \sum_{k=0}^{15} p_k S(\hat{\rho}_{E|\beta_k}),  
\label{eq:xue}
\end{split}
\end{equation}
where $S(\hat{\rho}_E)=-\mathrm{Tr}[\hat{\rho}_E\mathrm{log}_{2} \hat{\rho}_E]$ represents the von Neumann entropy associated with state $\hat{\rho}_E$,  and $S(\hat{\rho}_{E|\beta_k})$ is the von Neumann entropy of Eve's state when the integrated sender sends the $k$-th state $\beta_k$. For the pure state $|\varepsilon_{k}\rangle\langle\varepsilon_{k}|$, we have $S(\hat{\rho}_{E|\beta_k})=0$. 

If reverse reconciliation (RR) is applied, Eve's information about the dealer's state is bounded by the Holevo bound $\chi_{DE}$ as
\begin{equation}
\begin{split}
\chi\rightarrow\chi_{DE} = S(\hat{\rho}_E) - \sum_{k=0}^{15} P(k) S(\hat{\rho}_{E|\tilde{\beta}_k}),  
\label{eq:xde}
\end{split}
\end{equation}
where $P(k)=\frac{1}{15} \sum_{k=0}^{15} p(\beta_k|n_{M-1})$ is the overall probabilities of the dealer’s detection associated with outcome $\tilde{\beta}_k$, and $\hat{\rho}_{E|\tilde{\beta}_k}$ is the state of Eve when the dealer obtains the $k$-th result \cite{notarnicola2023optimizing}, and it can be expressed as 
\begin{equation}
\begin{split}
\begin{aligned}
\hat{\rho}_{E|\tilde{\beta}_k}=\frac{1}{16P(k)}\sum_{k=0}^{15}P(\beta_{k}|n_{M-1})|\varepsilon_{k}\rangle\langle\varepsilon_{k}|. \label{eq:rc}
\end{aligned}
\end{split}
\end{equation}
The von Neumann entropies of Eq. (\ref{eq:r}) and Eq. (\ref{eq:rc}) can be computed with the methods detailed in Appendix. \ref{app:3}. 

Finally, the secret key rate of SDD-CVQSS with DR can be calculated using Eq. (\ref{eq:IIUD}) and Eq. (\ref{eq:xue}), and the secret key rate of SDD-CVQSS with RR can be calculated using Eq. (\ref{eq:IIUD}) and Eq. (\ref{eq:xde}).

\section{PERFORMANCE ANALYSIS AND DISCUSSION}\label{V}

For now, the proposed SDD-CVQSS protocol, including its security model and secret key rate calculation, has been presented in detail. In what follows, we quantitatively evaluate the performance of the proposed SDD-CVQSS protocol.

Our numerical simulations consider noise and device defects as global influencing parameters, including detection efficiency, thermal noise, imperfect visible light interference, dark count noise, and transmittance of beam splitter. These global parameters are listed in Table \ref{tab:1} where their values are set according to the realistic experimental environment \cite{becerra2011m,becerra2013experimental,zhao2024security}. Unless specified otherwise, the following simulations are performed under the situation that user 2 is located at the midpoint between user 1 and the dealer.

\begin{table}[htbp]
\begin{ruledtabular}
\caption{Global parameters for numerical simulations}
\begin{tabular}{c c c}
\textrm{Symbol}&
\textrm{Value}&
\textrm{Description}\\
\colrule
$\eta_{s}$ & 0. 72 & SDD's detection efficiency\\
$\mathrm{N}_t$ & 0. 01 & Thermal noise\\
$\xi$ & 0. 998 & Phase noise \\
$\nu$ & 0. 001 & Dark count \\
$\zeta$ & 0. 99 & Transmittance of beam splitter \\
$\eta_{PHD}$ & 0. 72 & PHD's detection efficiency \\
$v_{el}$ & 0. 01 & PHD's electronic noise \\
\end{tabular}
\label{tab:1}
\end{ruledtabular}
\end{table}

\begin{figure}[htbp]
\includegraphics[width = \columnwidth]{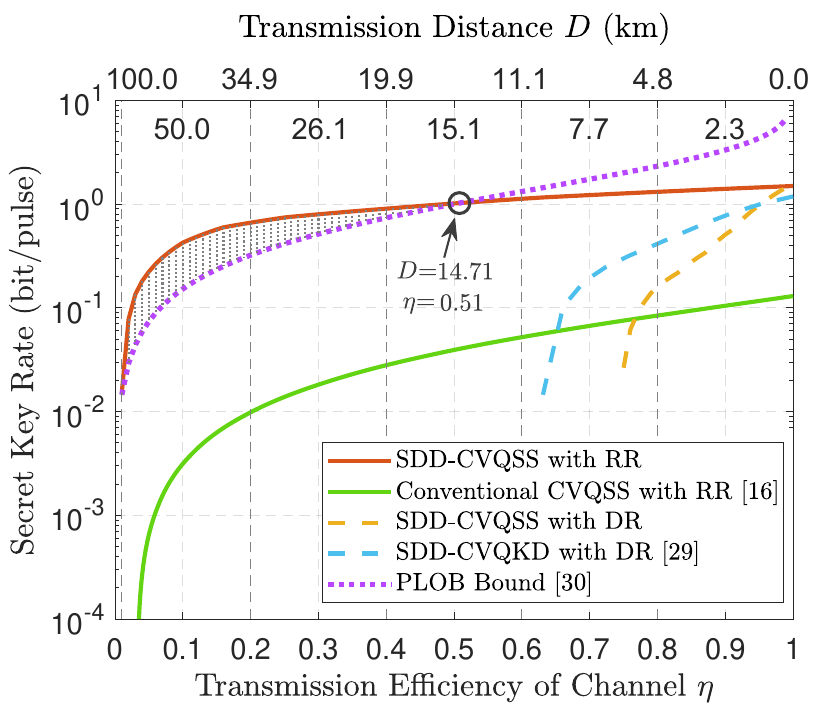}
\caption{Secret key rate of SDD-CVQSS as functions of transmission distance $D$ (between user 1 and the dealer) and transmission efficiency of the channel with $M=4$ adaptive measurements. The modulation variance is optimized within the range of $(0,20]$.}
\label{fig:skr}
\end{figure}

Fig. \ref{fig:skr} shows the asymptotic performance of SDD-CVQSS with four adaptive measurement rounds. For comparison, the asymptotic performance of conventional CVQSS \cite{liao2021quantum} and SDD-CVQKD \cite{zhao2024security} are also plotted. It can be found that the performance of our proposed SDD-CVQSS with RR (red solid line) significantly outperforms all other schemes in terms of both maximum transmission distance and secret key rate. This improvement is mainly attributed to two aspects, i.e., the data reconciliation strategy and the deployment of SDD. For the data reconciliation strategy, specifically, RR has been proven to outperform DR which is constricted by 3 dB limitation \cite{silberhorn2002continuous}. This advantage is why the maximum transmission distance of SDD-CVQSS with RR (red solid line) is superior to the DR version (yellow dashed line). For the deployment of SDD, its improvement in secret key rate is explicitly reflected by the comparison between the SDD-CVQSS with RR (red solid line) and the conventional CVQSS with RR (green solid line), thereby illustrating SDD's significant advantage in enhancing CVQSS secret key rate. Remarkably, the proposed SDD-CVQSS with RR even surpasses the PLOB bound (purple dotted line) when transmission distance reaches 14.71 km. This breakthrough stems from the operational mechanism of SDD, which enables accurate discrimination and precise recovery of the quantum state transmitted by the sender. Specifically, $|\widetilde{\beta}_{k}\rangle=|{\beta}_{k}'\rangle$ if the inference is correct. This is conceptually analogous to deploying a noiseless amplifier \cite{PhysRevA.86.012327} in front of the dealer's entrance to completely compensate for the negative effects caused by the channel loss and excess noise, providing a feasible approach for CVQSS to surpass the PLOB bound.

\begin{figure}[htbp]
\includegraphics[width = \columnwidth]{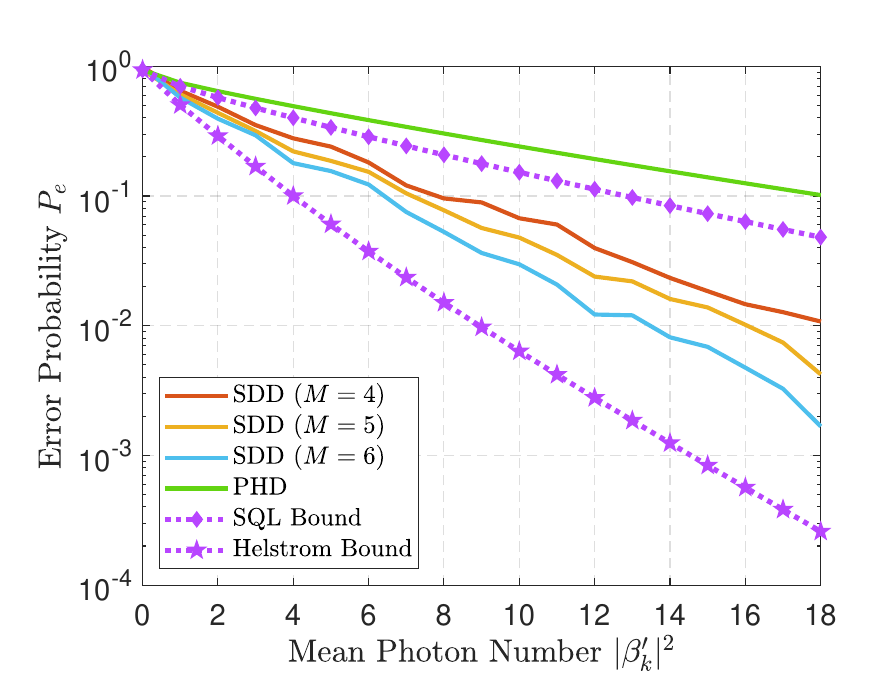}
\caption{Error probabilities of SDD as functions of mean photon number $|\beta_k'|^2$ for a transmission distance $D=50$ km. Purple dotted lines with different marks respectively denote the SQL bound and the Helstrom bound for discriminating 16 possible mixed states.}
\label{fig:Pe}
\end{figure}

To investigate why the secret key rate of CVQSS can be greatly improved by SDD, Fig. \ref{fig:Pe} illustrates the performance of SDD in terms of error probability, which characterizes a quantum detector's capability to correctly distinguish non-orthogonal coherent states. For comparative analysis, we also plot the error probabilities of practical heterodyne detector (PHD), the SQL bound and the Helstrom bound as functions of mean photon number. See Appendix \ref{app:2} for the calculations of these error probabilities. It can be easily found that the error probability of PHD (green solid line) is always higher than the SQL bound (purple diamond-dotted line) due to its non-ideal characteristics of practical devices. By contrast, the proposed SDD with four adaptive measurement rounds (red solid line) exhibits a lower error probability that consistently remains below the SQL bound. These quantitative results verify the conjecture that our proposed SDD is able to surpass the SQL bound, while PHD cannot. This implies that the proposed SDD can correctly distinguish 16 possible mixed states with much lower error probability, which is beneficial for enhancing the secret key rate of CVQSS. Moreover, it is further observed that the error probability of SDD decreases as the number of adaptive measurement rounds $M$ increases. This is because the amount of historical data accumulates with $M$, enabling SDD to derive a more precise prediction of the estimated state $|\widetilde{\beta}_{k}\rangle$ via Bayesian inference. Note that although the computational power limitations of simulation devices prevent us from obtaining the performance of SDD for larger values of $M$, based on existing results and the principle of SDD, it is believed that the error probability of SDD can theoretically approach or even reach the Helstrom bound (purple star-dotted line) when $M$ is sufficiently large.

\begin{figure}[htbp]
\includegraphics[width = \columnwidth]{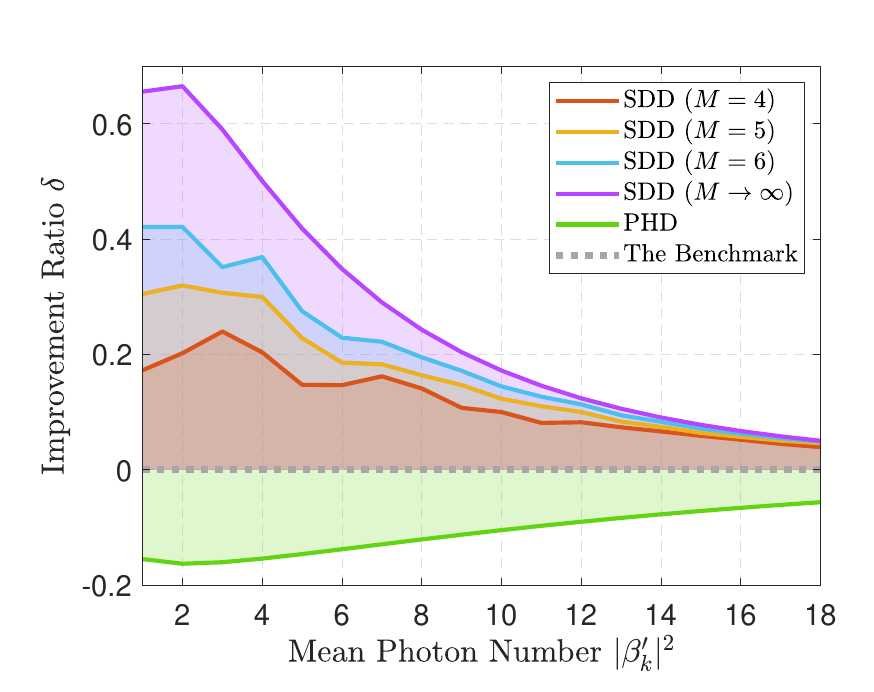}
\caption{Improvement ratio $\delta$ as functions of mean photon number $|\beta_k'|^2$ for a transmission distance $D=50$ km.}
\label{fig:Gs1}
\end{figure}

To visualize the magnitude of SDD's performance enhancement for CVQSS, an improvement ratio $\delta$ is defined by the formula
\begin{equation}
\begin{split}
\begin{aligned}
\delta=-\frac{P_e-P_{SQL}}{1-P_{SQL}}, \label{eq:ped}
\end{aligned}
\end{split}
\end{equation}
where the value of $P_e=\{P_e^{SDD}(M), P_e^{PHD}(\eta_{PHD},v_{el})\}$ is dependent on the detector employed. Evidently, this metric establishes the SQL bound as a benchmark for evaluating detectors’ performance, since $P_e=P_{SQL}$ when $\delta=0$. As shown in Fig. \ref{fig:Gs1}, although the improvement ratios of all SDDs (solid lines except green) decrease with increasing mean photon number, they consistently stay above the benchmark (gray dashed line), and the magnitude of improvement increases as $M$ grows. In contrast, the PHD (green solid line) exhibits a negative effect on the performance improvement as its value remains negative and below the benchmark. These results intuitively confirm the proposed SDD's effectiveness in surpassing the SQL bound, even under non-ideal conditions.

\begin{figure}[htbp]
\includegraphics[width = \columnwidth]{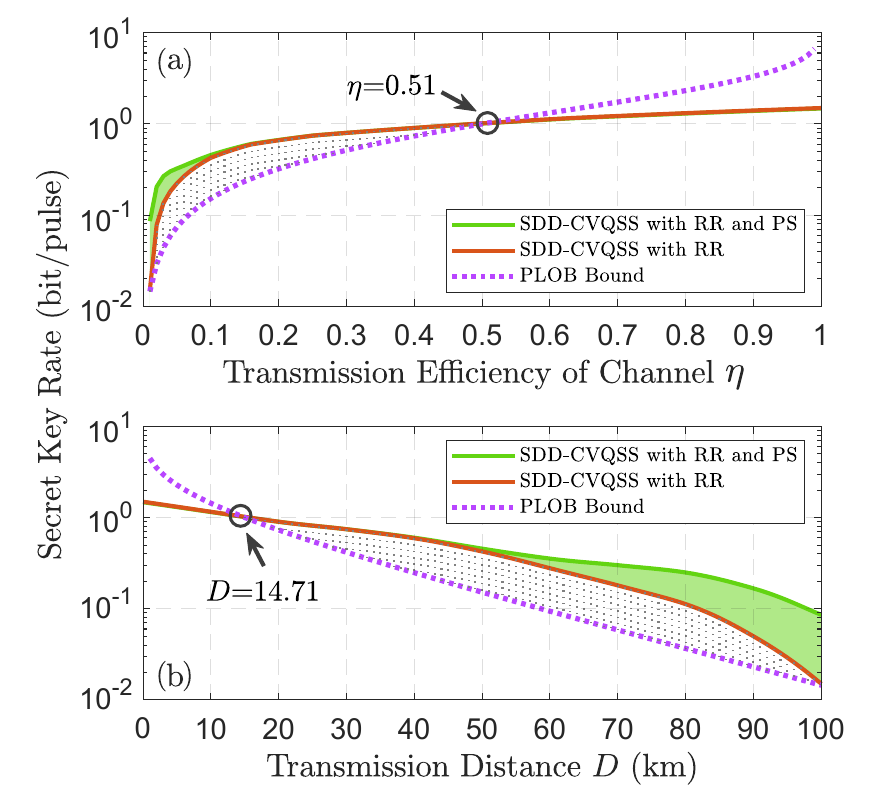}
\caption{Secret key rate of SDD-CVQSS with $M=4$ adaptive measurements as functions of (a) transmission efficiency of channel $\eta$ and (b) transmission distance $D$. The green shaded areas indicate the performance improvement gained from the PS scheme.}
\label{fig:tutu}
\end{figure}

In addition, the performance of SDD-CVQSS can be further enhanced using a post-selection (PS) scheme \cite{PhysRevA.73.052316}, which enables users and the dealer to retain only those events that they are closely correlated, thereby gaining an advantage over Eve. See Appendix \ref{app:4} for the derivation of the secret key rate of SDD-CVQSS with PS. Fig. \ref{fig:tutu}(a) depicts the performance improvement of SDD-CVQSS with PS as a function of channel transmission efficiency, while Fig. \ref{fig:tutu}(b) presents its performance improvement as a function of linearly varying transmission distance under the same conditions. As can be seen from both subfigures, the performance of SDD-CVQSS with RR can be explicitly enhanced using PS scheme, which demonstrates the effectiveness of PS in improving the secret key rate of the proposed SDD-CVQSS. Specifically, as shown in Fig. \ref{fig:tutu}(a), the performance improvement gained from the PS scheme can be observed (green shaded area in (a)) when the channel transmission efficiency $\eta$ is approximately less than 0.1. This is consistent with Fig. \ref{fig:tutu}(b) in which the performance improvement from the PS scheme becomes evident (green shaded area in (b)) when the transmission distance $D$ reaches approximately 50 km, since lower channel transmission efficiency tends to imply a longer transmission distance. 
Note that the PS scheme cannot substantially improve the performance of SDD-CVQSS with RR when $\eta$ exceeds 0.1 (i.e., $D$ is less than 50 km). This is because, within this range where channel loss is much smaller, almost all events in which users and the dealer are closely correlated always render $I_{UD}>\chi_{DE}$, such that the dealer rarely needs to discard measurement results. Therefore, the PS scheme is beneficial for improving the secret key rate of SDD-CVQSS in long-distance transmission scenarios.

\begin{figure}[htbp]
\includegraphics[width = \columnwidth]{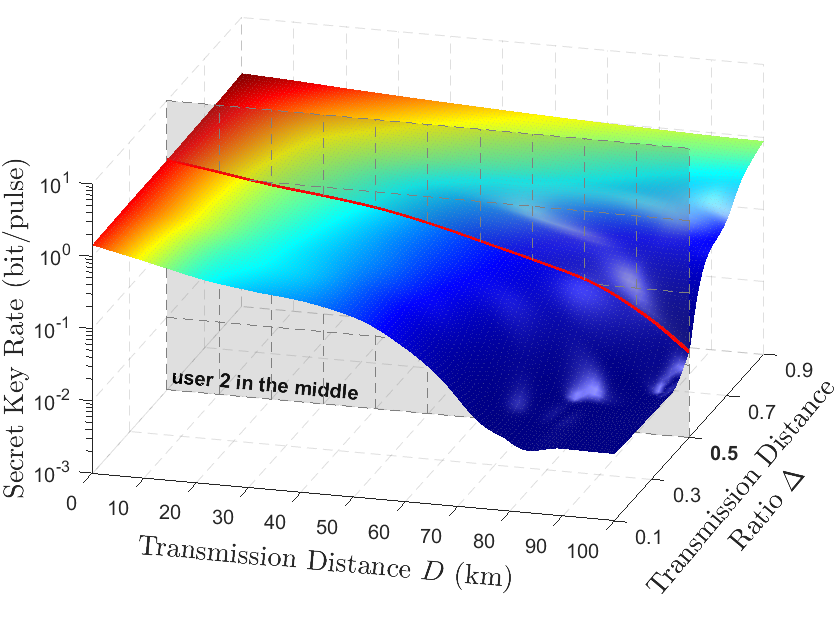}
\caption{Secret key rate of SDD-CVQSS with RR as functions of transmission distance $D$ and the transmission distance ratio $\Delta$. The red solid line represents the performance of SDD-CVQSS with RR when user 2 is positioned at the midpoint, which is exactly identical to the red solid line in Fig. \ref{fig:tutu}(b).}
\label{fig:unskr}
\end{figure}
\begin{figure}[htbp]
\includegraphics[width = \columnwidth]{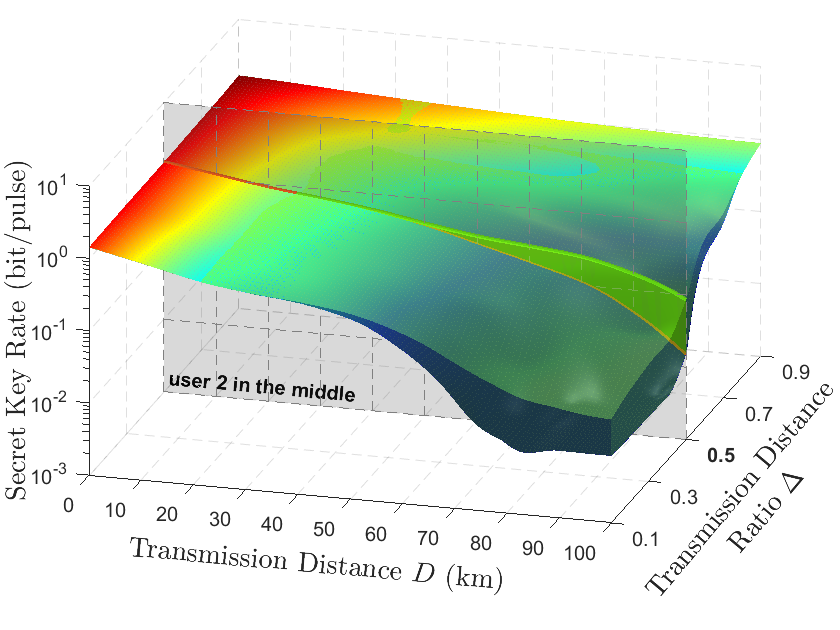}
\caption{Performance compensation for SDD-CVQSS using the PS scheme. The green semi-transparent 3D region denotes the improvement gained from the PS scheme, and the green solid line represents the performance of SDD-CVQSS with RR and PS when user 2 is positioned at the midpoint, which is exactly identical to the green solid line in Fig. \ref{fig:tutu}(b).}
\label{fig:unskr2}
\end{figure}
Up to now, all the performance analyzed above is based on a symmetric situation where user 2 is located at the midpoint between user 1 and the dealer, i.e., the transmission distances from user 1 to user 2 and from user 2 to the dealer are equal. In considering the practical implementation of SDD-CVQSS, however, user 2 should be able to be flexibly positioned according to his needs. To investigate the impact of the asymmetric situation on the performance of SDD-CVQSS, we define a transmission distance ratio $\Delta=d/D$ where $d$ denotes the distance from user 1 to user 2, and then plot Fig. \ref{fig:unskr}, which shows the secret key rate of SDD-CVQSS with RR as functions of the transmission distance $D$ and the transmission distance ratio $\Delta$. It can be observed that there exists an overall trend for all transmission distances where the closer user 2 is to user 1, the lower the secret key rate. This is because, from the dealer's perspective, some of the 16 mixed states such as $|\beta_1\rangle$ and $|\beta_4\rangle$ are quite similar when user 1 and user 2 are not far apart, making it difficult for the SDD to discriminate them correctly. This situation is more severe when discriminating $|\beta_2\rangle$, $|\beta_7\rangle$, $|\beta_8\rangle$ and $|\beta_{13}\rangle$, since they largely overlap near the origin of phase space due to the short distance $d$. That is to say, the error probability of the SDD in discriminating these overlapping states increases as user 2 gradually approaches user 1, resulting in performance degradation of SDD-CVQSS. It can be further observed that the trend of performance degradation in SDD-CVQSS becomes evident when the transmission distance $D$ reaches approximately 50 km. This indicates that user 2 should be positioned as far as possible from user 1 to achieve better performance in long-distance transmission scenarios. Fortunately, we can still enhance the performance of SDD-CVQSS using the PS scheme even if practical requirements prevent user 2 from being positioned far from user 1. Fig. \ref{fig:unskr2} demonstrates that the performance degradation of SDD-CVQSS, particularly in long-distance transmission scenarios, can be efficiently compensated by this scheme, offering a feasible solution to boost the secret key rate of SDD-CVQSS in such scenarios.

\section{CONCLUSION}\label{VI}
In this work, we have proposed a novel CVQSS protocol integrated with an SDD (SDD-CVQSS). We first developed the procedure of SDD-CVQSS, and then detailed the specially designed SDD. We subsequently constructed a security model for SDD-CVQSS and finally derived its security bound against beam-splitting collective attacks. The asymptotic performance of SDD-CVQSS was deeply analyzed, and the numerical results demonstrated that our proposed SDD-CVQSS with RR is able to surpass the PLOB bound, thereby significantly improving the performance of the CVQSS system. In addition, a PS scheme was suggested to compensate for the performance degradation of SDD-CVQSS in long-distance transmission scenarios, providing a feasible way to achieve high-performance CVQSS.

This work reveals the improvement mechanism of SDD-CVQSS, i.e., the more adaptive measurements the SDD conducts, the more precise the estimated state is predicted. This mechanism helps reduce the error probability of discriminating mixed states at the dealer's side, which is beneficial for enhancing the performance of the CVQSS system. In our future study, we will develop quantum algorithms \cite{liao2025high} to assist in estimating the secret key rate of SDD-CVQSS with more rounds of adaptive measurements.

\begin{acknowledgments}
This work was supported by the Hunan Provincial Key Research and Development Program (Grant No. 2025QK3011) and the National Natural Science Foundation of China (Grant No. 62101180).
\end{acknowledgments}

\appendix
\section{The derivation of the posterior probability for each possible state $|\beta_k\rangle$}\label{app:1}

For the first adaptive measurement, the posterior probability for a certain mixed state $|\beta_k\rangle$ can be calculated by Bayes' theorem \cite{becerra2013experimental}
\begin{align}
P(\beta_{k}|n_{0})=\frac{p_kP(n_{0}|\beta_{k},\gamma_{s_{0}})}{\sum_{k=0}^{15}p_kP(n_{0}|\beta_{k},\gamma_{s_{0}})}, 
\end{align}
where $p_k$ is the prior probability of the mixed state $|\beta_k\rangle$, $P(n_{0}|\beta_{k},\gamma_{s_{0}})$ is conditional probability of observing photon number $n_{0}$ for $|\beta_k\rangle$ given the LO field $\gamma_{s_{0}}$. This posterior probability is deemed the prior probability of next adaptive measurement, such that the posterior probability for mixed state $|\beta_k\rangle$ on 1-th branch can be expressed as
\begin{align}
P(\beta_{k}|n_0,n_{1})&=\frac{P(\beta_{k}|n_{0})P(n_{1}|\beta_{k},\gamma_{s_{1}})}{\sum_{k=0}^{15}P(\beta_{k}|n_{0})P(n_{1}|\beta_{k},\gamma_{s_{1}})}\nonumber\\&=\frac{p_kP(n_{0}|\beta_{k},\gamma_{s_{0}})P(n_{1}|\beta_{k},\gamma_{s_{1}})}{\sum_{k=0}^{15}p_kP(n_{0}|\beta_{k},\gamma_{s_{0}})P(n_{1}|\beta_{k},\gamma_{s_{1}})}. 
\end{align}

By iteratively applying Eq. (\ref{eq:post}) and simplifying the formula, the posterior probability of $|\beta_k\rangle$ in $(M-1)$-th branch can be calculated as 
\begin{align}
&P(\beta_{k}|n_0,...,n_{M-1}) \nonumber \\&=\frac{p_{k}P(n_{0}|\beta_{k}, \gamma_{s_{0}})\prod_{i=1}^{M-1}P(n_{i}|\beta_{k}, \gamma_{s_{i}})}{\sum_{k=0}^{15}p_{k}P(n_{0}|\beta_{k}, \gamma_{s_{0}})\prod_{i=1}^{M-1}P(n_{i}|\beta_{k}, \gamma_{s_{i}})}.
\end{align}

\section{The calculation of Von Neumann entropy}\label{app:3}
In general, Eq. (\ref{eq:r}) and Eq. (\ref{eq:rc}) have the following identical form

\begin{equation}
\begin{split}
\hat{\rho}=\sum_{k=0}^{15}c_k|\varepsilon_{k}\rangle\langle\varepsilon_{k}|. 
\label{eq:rho}
\end{split}
\end{equation}
To calculate the Von Neumann entropy, we need to diagonalize the state $\hat{\rho}$. Let the eigenvector of Eq. (\ref{eq:rho}) be $\mid\psi\rangle=\sum_{k=0}^{15}b_{k}\mid\varepsilon_{k}\rangle$, we have
\begin{equation}
\begin{split}
\lambda\mid\psi\rangle=\hat{\rho}\mid\psi\rangle, 
\end{split}
\end{equation}
namely,
\begin{equation}
\begin{split}
\lambda\sum_{s=0}^{15}b_{s}\mid\varepsilon_{s}\rangle&=\sum_{k=0}^{15}c_k|\varepsilon_{k}\rangle\langle\varepsilon_{k}|\sum_{m=0}^{15}b_{m}\mid\varepsilon_{m}\rangle \\&=\sum_{k=0}^{15}c_{k}(\sum_{m=0}^{15}G_{km}b_{m})\mid\varepsilon_{k}\rangle,  
\end{split}
\end{equation}

where 
\begin{equation}
\begin{split}
G_{km}&=\left\langle\beta_{k}\mid\beta_{m}\right\rangle = \exp\{-\frac{1}{2}|\beta_{k}-\beta_{m}|^{2}\} 
\end{split}
\end{equation}
denotes the overlap between $|\beta_{k}\rangle$ and $|\beta_{m}\rangle$.

Then
\begin{equation}
\begin{split}
\lambda &b_{k}=c_{k}\left(\sum_{m=0}^{15}G_{km}b_{m}\right), 
\end{split}
\end{equation}
or, equivalently,
\begin{equation}
\begin{split}
\left(\frac{\lambda}{c_{k}}-1\right)b_{k}-\sum_{m\neq k}G_{km}b_{k_m}=0. 
\end{split}
\end{equation}

The above formula can be rewritten as $W = 0$, $W$ is given by the following matrix

\begin{equation}
\begin{split}
W=\begin{pmatrix}\frac{\lambda}{c_0}-1&-G_{01}&\cdots&-G_{0\ 15}\\\\-G_{10}&\ddots&\ddots&\vdots\\\\\vdots&\ddots&\ddots&-G_{14\ 15}\\\\-G_{15\ 0}&\cdots&-G_{15\ 14}&\frac{\lambda}{c_{15}}-1\end{pmatrix}. 
\end{split}
\end{equation}

The eigenvalues $\lambda_k$ of the calculated quantum state can be solved by calculating the determinant of the $W$ matrix to be 0 and $det\ W = 0$. This provides us with the eigenvalues $\lambda_k$ and the corresponding
von Neumann entropy $S[\rho]=-\sum_{k=0}^{15}\lambda_k\log_2\lambda_k$.

\section{The calculation of error probabilities}\label{app:2}

Here we derive the error probabilities for the proposed SDD, as well as the SQL bound and the Helstrom bound, with a 16QAM-like modulation strategy.

For SDD, $|\tilde{\beta}\rangle$ is predicted as the final detection result, and the detection history sequence $\{n_0,...,n_{M-1}\}$ denotes the number of photons observed in each adaptive measurement. The probability of correct decision is given by $P(\tilde{\beta}|n_0,...,n_{M-1})$, and the error probability can be expressed as the complement of this probability, i.e.,
\begin{equation}
P_{e}^{SDD}=1-\sum_{n_{0}=0}^{\infty}... \sum_{n_{M-1}=0}^{\infty}P(n_0,...,n_{M-1})P(\tilde{\beta}|n_0,...,n_{M-1}), 
\end{equation}
where $P(n_0,...,n_{M-1})=\sum_{k=0}^{15} p_k P(n_0, \ldots, n_{M-1} | \beta_k)$ represents the probability of detecting $\{n_0,...,n_{M-1}\}$ given the input signal $\beta_k$.

It is evident that $P(\tilde{\beta}|n_0,...,n_{M-1})=p_{|\tilde{\beta}\rangle}P(n_0,...,n_{M-1}|\tilde{\beta})/P(n_0,...,n_{M-1})$. Since $P(n_0,...,n_{M-1}|\tilde{\beta})=P(n_{0}|\tilde{\beta}, \gamma_{s_{0}})\prod_{i=1}^{M-1}P(n_{i}|\tilde{\beta}, \gamma_{s_{i}})$, the error probability of SDD can be written as
\begin{align}
&P_{e}^{SDD}=1-\sum_{n_{0}=0}^{\infty}... \sum_{n_{M-1}=0}^{\infty}p_{|\tilde{\beta}\rangle}P(n_0,...,n_{M-1}|\tilde{\beta}) \nonumber\\& =1-\sum_{n_{0}=0}^{\infty}...\sum_{n_{M-1}=0}^{\infty}p_{|\tilde{\beta}\rangle}P(n_{0}|\tilde{\beta}, \gamma_{s_{0}})\prod_{i=1}^{M-1}P(n_{i}|\tilde{\beta}, \gamma_{s_{i}}). 
\end{align}

\begin{figure}[htbp]
\includegraphics[width = \columnwidth]{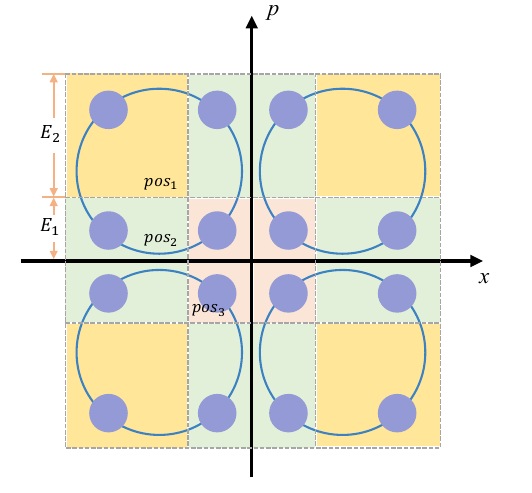}
\caption{The constellation diagram of the 16QAM-like modulation strategy in phase space. Three distinct regions represent different types of states: corner points (yellow region), edge points (green region), and internal points (pink region) are defined as $pos_1$, $pos_2$ and $pos_3$, respectively.}
\label{fig:ee}
\end{figure}
The SQL arises from inherent quantum uncertainty, such that even an ideal coherent detector with perfect efficiency exhibits an error probability when discriminating among mixed states. This error probability can be expressed as
\begin{equation}
\begin{split}
P_{e}^{SQL}=\sum_{k=0}^{15}p_k\cdot P_{pos},
\end{split}
\end{equation}
where $P_{pos}=\{P_{pos_1},P_{pos_2},P_{pos_3}\}$ denotes the error probabilities of distinct regions as shown in Fig.\ref{fig:ee}. The error probabilities of different regions depend on the probability of being mismeasured as an adjacent mixed state during the measurement of the region \cite{Proakis2007Digital}, given by
\begin{equation}
\begin{split}
P_{pos_1}=2P_{E_1}-P_{E_1}^2,
\end{split}
\end{equation}
\begin{equation}
\begin{split}
P_{pos_2}=2P_{E_1}+P_{E_2}-(P_{E_1}+P_{E_2})\times P_{E_1},
\end{split}
\end{equation}
\begin{equation}
\begin{split}
P_{pos_3}=2(P_{E_1}+P_{E_2})-(P_{E_1}+P_{E_2})^2,
\end{split}
\end{equation}
where $E_1$, $E_2$ represent the two Euclidean distances of adjacent mixed states, and $P_{E}$, which represents the probability of error judgment under these two Euclidean distances, is calculated as 
\begin{equation}
\begin{split}
P_{E_i}=\frac{1}{2} \mathrm{erfc}(\frac{E_i}{2})\;\;\;\;i=\{1,2\},
\end{split}
\end{equation}
where the Gaussian error function
\begin{equation}
\begin{split}
\mathrm{erfc}(x)=\frac{2}{\sqrt{\pi}}\int_x^\infty e^{-t^2}\mathrm{d}t.
\end{split}
\end{equation}

The Helstrom bound is the theoretical lower bound of the minimum error probability in quantum state discrimination, which can be approximated by the square-root measure \cite{liao2018long} and expressed as
\begin{equation}
\begin{split}
P_{e}^{Hel}=1-\frac{1}{256}(\sum_{k=1}^{15}\sqrt{\omega_k})^2,
\end{split}
\end{equation}
where $\omega_k$ are the eigenvalues of the Gram matrix $G$ for the mixed states, defined as
\begin{equation}
\begin{split}
G=\Bigg(\langle\beta_{k}\mid\beta_{m}\rangle\Bigg)_{k,m=0,...,15},
\end{split}
\end{equation}
with $\langle\beta_{k}\mid\beta_{m}\rangle$ denoting the inner product of quantum states $|\beta_{k}\rangle$ and $|\beta_{m}\rangle$.

\section{The derivation of the asymptotic secret key rate for SDD-CVQSS with PS}\label{app:4}

The PS scheme enables the dealer to first select the measurement results where it holds an advantage over Eve and discard the rest after receiving and measuring the mixed states sent by the integrated sender. This is able to partially compensate for the loss of RR with an imperfect detector \cite{PhysRevA.73.052316}. Specifically, the dealer calculated the mutual information $I_{UD}(n_0,...,n_{M-1})$ after he obtains the photon number result $(n_0,...,n_{M-1})$ of each round adaptive measurement. The measurement results with negative secret key rate are discarded, i.e., only those satisfying $I_{UD}(n_0,...,n_{M-1}) >\chi_{DE}$ are retained. 

The mutual information $I_{UD}(n_0,...,n_{M-1})$ for certain measurement result $(n_0,...,n_{M-1})$ can be calculated by 
\begin{equation}
\begin{split}
&I_{UD}(n_0,...,n_{M-1})=H(U)-H(U|n_0,...,n_{M-1})\\&=-\sum_{k=0}^{15}p_{k}\mathrm{log}_{2}p_{k}+\sum_{k=0}^{15}P(\beta_{k}|n_{0}, ... , n_{M-1}) \\&\times\mathrm{log}_{2}P(\beta_{k}|n_{0}, ... , n_{M-1}).
\label{eq:iudn}
\end{split}
\end{equation}

Combining formulas Eq. (\ref{eq:est_beta}) and Eq. (\ref{eq:iudn}), the measured mutual information $I_{UD}(n_0,...,n_{M-1})$ for a specific result $(n_0,...,n_{M-1})$ is

\begin{widetext}
\begin{equation}
\begin{split}
&I_{UD}(n_0,...,n_{M-1})=H(U)-H(U|n_0,...,n_{M-1})\\&=-\sum_{k=0}^{15}p_{k}\left(\log_{2}p_k-\frac{P(n_{0}|\beta_{k}, \gamma_{s_{0}})\prod_{i=1}^{M-1}P(n_{i}|\beta_{k}, \gamma_{s_{i}})}{\sum_{k=0}^{15}p_{k}P(n_{0}|\beta_{k}, \gamma_{s_{0}})\prod_{i=1}^{M-1}P(n_{i}|\beta_{k}, \gamma_{s_{i}})}   \log_{2}\frac{p_{k}P(n_{0}|\beta_{k}, \gamma_{s_{0}})\prod_{i=1}^{M-1}P(n_{i}|\beta_{k}, \gamma_{s_{i}})}{\sum_{k=0}^{15}p_{k}P(n_{0}|\beta_{k}, \gamma_{s_{0}})\prod_{i=1}^{M-1}P(n_{i}|\beta_{k}, \gamma_{s_{i}})}\right). 
\label{eq:iudn1}
\end{split}
\end{equation}
\end{widetext}

Then the asymptotic secret key rate for SDD-CVQSS with PS can be derived as
\begin{equation}
\begin{split}
R_{PS}=\sum_{p\in S}p(n_0,...,p_{M-1})(I_{UD}(n_0,...,n_{M-1})-\chi_{DE}),
\label{eq:rps}
\end{split}
\end{equation}
where $S=\{p|I_{UD}(n_0,...,n_{M-1})-\chi_{DE}>0\}$ consists of measurement results $(n_0,...,n_{M-1})$ that enables a positive secret key rate.


\nocite{*}
\bibliography{REF1}
\end{document}